\documentclass[twocolumn,prd,amsmath,amssymb,floatfix]{revtex4}

\usepackage{bbold}
\usepackage{bm}
\usepackage{color}
\usepackage{dcolumn}
\usepackage{feynmf} 
\usepackage{graphics}
\usepackage{graphicx}
\usepackage{subfigure}
\usepackage{slashed}
\usepackage{placeins}

\allowdisplaybreaks
\usepackage{pdfpages}
\usepackage{eso-pic}
\usepackage{everyshi}
\usepackage{pdflscape}

\usepackage{epsfig}

\def\al{\alpha}

\def\be{\begin{equation}}
\def\ee{\end{equation}}
\def\bea{\begin{eqnarray}}
\def\eea{\end{eqnarray}}
\def\bse{\begin{subequations}}
\def\ese{\end{subequations}}
\def\bc{\begin{center}}
\def\ec{\end{center}}

\def\nonum{\nonumber}

\begin{document}

\title{Addendum to ``Critical behaviour of ($2+1$)-dimensional QED:\\
       $1/N_f$-corrections in an arbitrary non-local gauge''}
       \author{
       A.V.~Kotikov$^{1}$ and S.~Teber$^{2}$}
\affiliation{
$^1$Bogoliubov Laboratory of Theoretical Physics, Joint Institute for Nuclear Research, 141980 Dubna, Russia.\\
$^2$Sorbonne Universit\'e, CNRS, Laboratoire de Physique Th\'eorique et Hautes Energies, LPTHE, F-75005 Paris, France.
 }

\date{\today}

\begin{abstract}
Dynamical chiral symmetry breaking (D$\chi$SB) is studied within ($2+1$)-dimensional QED
with $N$ four-component fermions. The leading and next-to-leading orders of the $1/N$ expansion
were computed exactly in Refs.~[\onlinecite{Gusynin:2016som,Kotikov:2016prf}]
in an arbitrary non-local gauge. In this addendum to [\onlinecite{Kotikov:2016prf}], we show that the resummation of the wave-function renormalization constant
at the level of the gap equation yields a {\it complete} cancellation
of the gauge dependence of the critical fermion flavour number resulting in: $N_c= 2.8469$,
which is such that D$\chi$SB takes place for $N<N_c$.
The result is in full agreement with one of Ref.~[\onlinecite{Gusynin:2016som}].
\end{abstract}

\maketitle

\section{Introduction }

We consider Quantum Electrodynamics in $2+1$ dimensions (QED$_3$) which is described by the Lagrangian:
\be
L = \overline \Psi ( i \hat \partial - e \hat A ) \Psi - \frac{1}{4} F_{ \mu \nu}^2\, ,
\label{L-QED3}
\ee
where $ \Psi$ is taken to be a four component complex spinor.
In the presence of $N$ fermion flavours, the model has a $U(2N)$ symmetry. A
fermion mass term, $m\overline \Psi \Psi$, breaks this symmetry to $U(N) \times U(N)$.
In a $1/N$ expansion \cite{AppelquistP81,JackiwT81+AppelquistH81},
the theory is super-renormalizable and the mass scale is then given by the dimensionful coupling constant: $a = Ne^2/8$, which is kept fixed as $N \rightarrow \infty$.

A central issue is related to the value of the critical fermion number, $N_c$, which is such that
D$\chi$SB
takes place only for $N<N_c$.
An accurate determination of $N_c$ is of crucial importance to understand the phase structure of QED$_3$.

In our studies Refs.~[\onlinecite{Kotikov:2016prf,KotikovST16}], we followed
the approach of Appelquist et al.~\cite{AppelquistNW88} who found that $N_c = 32/ \pi^2 \approx 3.24$ by solving
the Schwinger-Dyson (SD) gap equation in the Landau gauge using a leading order (LO) $1/N$-expansion.
Soon after the analysis of Ref.~[\onlinecite{AppelquistNW88}], Nash approximately included next-to-leading order (NLO) corrections and performed a partial resummation
of the wave-function renormalization constant at the level of the gap equation; he found [\onlinecite{Nash89}]: $N_c \approx 3.28$. Recently, upon refining the work of [\onlinecite{Kotikov93+12}],
the NLO corrections could be computed exactly in the Landau gauge yielding (in the absence of resummation) \cite{KotikovST16}: $N_c \approx 3.29$.
More recently, the results of Ref.~[\onlinecite{KotikovST16}] have been extended in Ref.~[\onlinecite{Kotikov:2016prf}] to an arbitrary non-local gauge
[\onlinecite{Simmons90+KugoM92}]. Ref.~[\onlinecite{Kotikov:2016prf}] then found a residual weak gauge-dependence of $N_c$ even after Nash's resummation; it was also noticed in
Ref.~[\onlinecite{Kotikov:2016prf}] that, if the weak gauge-dependent terms contributing to $N_c$ were neglected, then the final result would be in perfect agreement with
the one of Ref.~[\onlinecite{Gusynin:2016som}].

The purpose of this short note
is to upgrade
the exact results of [\onlinecite{Kotikov:2016prf}]
and to show the {\it complete} gauge-independence of the critical value $N_c$ in the $1/N^2$ approximation.
Following Ref.~[\onlinecite{Kotikov:2016prf}] and after long discussions with Valery Gusynin, we
shall modify 
 the expansion prescription used in Ref.~[\onlinecite{Kotikov:2016prf}] which was based on (an NLO correction to) the gap equation to (an NLO correction to) the parameter $\alpha$ of
its solution (see Eq.~(\ref{sigma-parametrization}) and below it). This subtle change in the interpretation of the NLO corrections
does not affect at all the LO results of Appelquist but significantly modifies the NLO results (see below Section~3)
leading to gauge-invariant $N_c$ values
after Nash's resummation.


\section{
Leading Order}
\label{sec:LO}

Let's briefly recall the structure and solutions of the LO SD equations, see [\onlinecite{Kotikov:2016prf}] for more details.
In the LO approximation to the $1/N$ expansion, the SD equation to the fermion propagator has the following form:
\begin{flalign}
  \Sigma (p) &= \frac{8(2+\xi)a}{N}
   \int \frac{d^3 k}{(2 \pi )^3}
\frac{ \Sigma (k) }{ \left( k^2 + \Sigma^2(k)
\right)
\bigl[ (p-k)^2 + a \,|p-k| \bigr]}
\nonum \\
&+ O(N^{-2})\, ,
\label{SD-sigma-LO1}
\end{flalign}
where $\Sigma (p)$ is the dynamically generated parity-conserving mass.

%
Following Ref.~[\onlinecite{Kotikov93+12}] and [\onlinecite{AppelquistNW88}], we consider the limit of large $a$ and linearize Eq.~(\ref{SD-sigma-LO1})
which yields:
\be
  \Sigma (p) =
 \frac{8(2 + \xi)}{N}
   \int \frac{d^3 k}{(2 \pi )^3} \frac{ \Sigma (k) }{k^2 \, |p-k| } + O(N^{-2})\, .
\label{SD-sigma-LO4}
\ee
The mass function may then be parametrized as \cite{AppelquistNW88}:
\be
\Sigma (k) = B \, (k^2)^{ -\alpha} \, ,
\label{sigma-parametrization}
\ee
where $B$ is arbitrary and the index $\al$ has to be self-consistently determined.
Using this ansatz, Eq.~(\ref{SD-sigma-LO4}) reads:
\be
\Sigma^{(\rm{LO})}(p) = \frac{4(2+\xi)B}{N}\,\frac{(p^2)^{-\al}}{(4\pi)^{3/2}}\, \frac{2\beta}{\pi^{1/2}}+ O(N^{-2})\, ,
\label{sigma-LO-res}
\ee
from which the LO gap equation is obtained:
\be
1 = \frac{(2+\xi)\beta}{L}+ O(L^{-2}) \quad  {\rm{or}} \quad \beta^{-1} = \frac{(2+\xi)}{L}+ O(L^{-2})\, , 
\label{gap-eqn-LO}
\ee
where
\be
 \beta = \frac{1}{\alpha \left( 1/2 - \alpha \right)} ~~ {\rm{and}} ~~  L \equiv \pi^2 N\, .
\label{parameters}
\ee
Let's note that the two equations in (\ref{gap-eqn-LO}) are completely equal to each other.
%
%
%
Solving the gap equation, yields:
\begin{eqnarray}
\alpha_{\pm} = \frac{1}{4}\,\left( 1 \pm \sqrt{1 - \frac{16(2+\xi)}{L}} \right) \, ,
\label{al-LO}
\end{eqnarray}
which reproduces the solution given by Appelquist et al.~\cite{AppelquistNW88}.
The gauge-dependent critical number of fermions: $N_c \equiv N_c(\xi) = 16(2+\xi)/ \pi^2$,
is such that $\Sigma(p) = 0$ for $N>N_c$ and:
\be
\Sigma(0) \simeq \exp \bigl[ - 2 \pi / (N_c/N - 1)^{1/2} \bigr]\, ,
\ee
for $N<N_c$. Thus, D$\chi$SB occurs when $\alpha$ becomes complex, that is for $N<N_c$.

\section{Next-to-leading order}
\label{sec:NLO}

Evaluating the NLO corrections to the SD-equation (\ref{SD-sigma-LO1}) yields (see Ref.~[\onlinecite{Kotikov:2016prf}])
the following gap equation:
\begin{flalign}
&1 = \frac{(2+\xi)\beta}{L} + \frac{1}{L^2}\,
\Bigl[8 S(\al,\xi) - 2(2+\xi) \hat{\Pi} \beta \Bigl .
\nonum \\
&\Bigr . + \left(-\frac{5}{3}+ \frac{26}{3}\xi -3\xi^2\right) \beta^2 - 8\beta\left(\frac{2}{3}(1-\xi)-\xi^2\right) \Bigr]
\nonum \\
& + O(L^{-3}) \, , \qquad
\label{gap-eqn-NLO-explicit}
\end{flalign}
where
\be
\hat{\Pi} = \frac{92}{9} - \pi^2\, ,
\label{sigma-NLO-1}
\ee
arises from the two-loop polarization operator in dimension $D=3$~[\onlinecite{Gracey93,Teber12+KotikovT13}]
~\footnote{Notice that $\hat{\Pi}$ has also been evaluated in Ref.~[\onlinecite{Kotikov93+12}] but it was not 
explicitly indicated in the corresponding NLO corrections.}.

The factor $S(\al,\xi)$ contains the contribution of the most complicated diagrams.
As it was shown in [\onlinecite{Kotikov:2016prf}], it is convenient to extract the most important contributions
$\sim \beta$ and $\sim \beta^2$ from the complicated part $S(\al,\xi)$. After 
theses calculations, 
the gap equation takes the equivalent form:
\begin{flalign}
&1 = \frac{(2+\xi)\beta}{L} + \frac{1}{L^2}
\Bigl[8 \tilde{S}(\al,\xi)
-2(2+\xi) \hat{\Pi} \beta  \Bigr .
\label{gap-eqn-NLO-explicit_1} \\
&\Bigl . + \left(\frac{2}{3}-\xi\right)\bigl(2+\xi\bigr)\, \beta^2 +
4\beta\left(\xi^2-\frac{4}{3}\xi - \frac{16}{3}\right) \Bigr] + O(L^{-3})\, , \quad
\nonum
\end{flalign}
%
%
where the new complicated part  $\tilde{S}(\al,\xi)$ does not contain any positive $\beta$ powers and
can be expanded in series of $\al^n$ (and, hence,
$\beta^{-n}$) starting with $n=0$.

\subsection{Gap equation}

 In Ref.~[\onlinecite{Kotikov:2016prf}] we have analyzed  Eq.~(\ref{gap-eqn-NLO-explicit}) at the critical point $\beta=16$ and found the corresponding critical value $L_c$.
 The same results can also be obtained from Eq.~(\ref{gap-eqn-NLO-explicit_1}).

 Here we will follow another strategy. As was already discussed in the Introduction, 
we will proceed in computing the NLO correction to the parameter $\beta^{-1}$ of the solution of the SD equation.
From (\ref{gap-eqn-NLO-explicit_1}), we have:
\begin{flalign}
&\beta^{-1} = \frac{2+\xi}{L} + \frac{1}{L^2}
\Bigl[\frac{8}{\beta} \tilde{S}(\beta,\xi) -2(2+\xi) \hat{\Pi}   \Bigr .
\label{beta.1} \\
&\Bigl . + \left(\frac{2}{3}-\xi\right)\bigl(2+\xi\bigr)\, \beta +
4\left(\xi^2-\frac{4}{3}\xi - \frac{16}{3}\right) \Bigr] + O(L^{-3})\, .
\nonum 
\end{flalign}
From this equation, it is clear that the first term in brackets 
is of the order of $\sim 1/L$ (as can be seen by solving Eq.~(\ref{beta.1}) iteratively) and thus its contribution is of the order of
$\sim 1/L^3$ and should therefore be neglected in the present analysis. So, with NLO accuracy, we 
obtain that:
\begin{flalign}
&\beta^{-1} = \frac{2+\xi}{L} + \frac{1}{L^2}
\Bigl[ \left(\frac{2}{3}-\xi\right)\bigl(2+\xi\bigr)\, \beta  -2(2+\xi) \hat{\Pi}   \Bigr .
\nonum \\
&\Bigl . +
4\left(\xi^2-\frac{4}{3}\xi - \frac{16}{3}\right) \Bigr] + O(L^{-3})\, .
 \label{beta.2}
\end{flalign}
We are now in a position to
compute $\beta^{-1}$ from Eq.~(\ref{beta.2}) as a combination of terms $\sim 1/L$ and $\sim 1/L^2$.
This is however not so important in the present analysis.
 Since
we are interested in the critical regime, we may derive $L_c$ in a straightforward way from (\ref{beta.2}) (or equally from Eq.~(\ref{gap-eqn-NLO-explicit_1}) with
the condition $ \tilde{S}(\beta,\xi)=0$) by setting $\beta=16$ and keeping the terms $O(1/L^2)$.
This yields:
\be
L_c^2 -16(2+\xi) L_c +32\left [(2+\xi) \hat{\Pi} + 2\xi \left(\frac{20}{3} + 3 \xi \right) \right ] = 0 \, .
 \label{Lc-eqn}
\ee
%
Solving Eq.~(\ref{Lc-eqn}), we have two standard solutions:
\begin{subequations}
\begin{flalign}
	&L_{c,\pm} = 8\left(2+\xi \pm \sqrt{d_1(\xi)}\right)\, ,
 \label{Lc} \\
	&d_1(\xi)= 4-\frac{8}{3}\xi-2\xi^2 - \frac{2+\xi}{2} \hat{\Pi}  \, .
\label{d1}
\end{flalign}
\label{Lc-solutions}
\end{subequations}
%
%
Combining these values with the one of $\hat{\Pi}$ in Eq.~(\ref{sigma-NLO-1}), yields:
\be
N_c(\xi=0)=3.17, \quad N_c(\xi=2/3)=2.91\, ,
\label{Nc-no-resum}
\ee
%
%
where ``$-$'' solutions are unphysical and there is no solution in the Feynman gauge ($\xi=1$).
The range of $\xi$-values for which there is a solution corresponds to $\xi_- \leq  \xi \leq  \xi_+$,
where $\xi_{+}=0.82$ and  $\xi_{-}=-2.24$. 
%
%

\subsection{Resummation}

Performing Nash's resummation, the gap equation takes the following form (see Ref.~[\onlinecite{Kotikov:2016prf}]):
\be
1 = \frac{8\beta}{3L}  + \frac{1}{L^2}\,\Bigl[ 8 \tilde{S}(\al,\xi)
- \frac{16}{3} \, \beta \, \left(\frac{40}{9} + \hat{\Pi} \right) \Bigr] + O(L^{-3})\, ,
\label{gap-eqn-NLO-explicit_5b}
\ee
%
which displays a strong suppression of the gauge dependence
as $\xi$-dependent terms do exist but they enter the gap equation only through the rest, $\tilde{S}$, which is very small numerically.

In Ref.~[\onlinecite{Kotikov:2016prf}] we have analyzed  Eq. (\ref{gap-eqn-NLO-explicit_5b}) at the critical point $\beta=16$ and found the corresponding critical value $L_c$.
By analogy with the previous subsection, we now proceed on finding the NLO correction to the parameter $\beta^{-1}$ of the solution of the SD equation. From (\ref{gap-eqn-NLO-explicit_5b}), this yields:
\be
\beta^{-1} = \frac{8}{3L}  + \frac{1}{L^2}\,\Bigl[ \frac{8}{\beta} \tilde{S}(\al,\xi) - \frac{16}{3}  \, \left(\frac{40}{9} +
\hat{\Pi} \right) \Bigr] + O(L^{-3})\, .
\label{beta.3}
\ee
From this equation, it is again clear that the first term in brackets is of the order of $\sim 1/L$ (as can be seen by solving Eq.~(\ref{beta.3}) iteratively)
and thus its contribution is $\sim 1/L^3$ and should be neglected in the present  analysis. So, we have:
\be
\beta^{-1} = \frac{8}{3L}  - \frac{1}{L^2}\,\frac{16}{3}  \, \left(\frac{40}{9} +
\hat{\Pi} \right) + O(L^{-3})\, ,
\label{beta.4}
\ee
which is now completely gauge-independent.

We now consider Eq.~(\ref{beta.4}) (or, equivalently, Eq.~(\ref{gap-eqn-NLO-explicit_5b}) with the condition $\tilde{S}(\beta,\xi)=0$) at the critical point $\al=1/4$ ($\beta=16$)
keeping all terms $O(1/L^2)$. This yields:
\bea
L_c^2 -\frac{128}{3} L_c  +\frac{256}{3}\,\left(\frac{40}{9} + \hat{\Pi} \right)  = 0 \, .
\label{Lc-eqn_1}
\eea
%
Solving Eq.~(\ref{Lc-eqn_1}), we have two standard solutions:
\begin{subequations}
\begin{flalign}
	&L_{c,\pm} = \frac{64}{3} \left(1 \pm \sqrt{d_2(\xi)}\right) \, ,
\label{Lc-2} \\
	&d_2(\xi)= 1-
 \frac{3}{16}
\,\left(\frac{40}{9} + \hat{\Pi} \right) =  \frac{1}{6} - \frac{3}{16} \, \hat{\Pi} \, ,
\label{dc-2}
\end{flalign}
\label{d2}
\end{subequations}
and we have for the ``$+$'' solution (the ``$-$'' one is nonphysical):
\be
\overline{L}_c=28.0981, \qquad \overline{N}_c=2.85\, .
\label{overlN}
\ee
The results of Eq.~(\ref{overlN}) are in full agreement with the recent results of [\onlinecite{Gusynin:2016som}].

\section{Conclusion }
\label{sec:conclusion}

We have studied D$\chi$SB in QED$_3$ by including $1/N^2$ corrections to the SD equation exactly and taking into account the full
$\xi$-dependence of the gap equation. Following Nash, the wave function renormalization constant has been resummed at the level of the gap equation
leading to a very weak gauge-variance of the critical fermion number $N_c$.


Reconsidering the NLO expansion of Ref.~[\onlinecite{Kotikov:2016prf}], we have implemented an NLO expansion for
the parameter $\beta^{-1}$ which is related to the index parametrising the mass-function rather than the mass function itself.
This prescription allowed us to show that the complicated weakly gauge-variant terms are actually of the order of $1/N^3$ and should be neglected in the present NLO analysis.
Thus, the obtained value $N_c=2.85$ is completely gauge independent and in full agreement with the one of Ref.~[\onlinecite{Gusynin:2016som}].
Both works [\onlinecite{Gusynin:2016som}] and [\onlinecite{Kotikov:2016prf}] are therefore in perfect agreement and yield order by order fully gauge-invariant methods to compute $N_c$.

\acknowledgments
We thank Valery Gusynin for illuminating discussions.
One of us (A.V.K.) was supported by RFBR grant 16-02-00790-a.

{\it Note added.} After this work was accepted for publication, we became aware of the papers [\onlinecite{Benvenuti:2018cwd,Li:2018lyb}] whose contents
partially overlap with ours.



\end{document}